\documentclass[aip,apl,groupaddress,showkeys,twocolumn,a4paper]{revtex4}
\usepackage{graphicx,soul}
\usepackage[left=2.0cm,top=2.0cm,right=2.0cm,bottom=2.0cm]{geometry}
\begin{document}

\title{Distinguishing impurity concentrations in GaAs and AlGaAs, using very shallow undoped heterostructures}

\author{W.Y. Mak}\author{K. Das Gupta}\email{kd241@cam.ac.uk} \author{H. E. Beere} \author{I. Farrer} \author{F. Sfigakis} \author{D. A. Ritchie}
\affiliation{Cavendish Laboratory, University of Cambridge, J.J. Thomson Avenue, Cambridge CB3 0HE, UK.}

\begin{abstract}
We demonstrate a method of making a very shallow, gateable, undoped 2-dimensional electron gas.
We have developed a method of making very low resistivity contacts
to these structures and  systematically  studied the evolution of the mobility as a function of the
depth of the 2DEG (from 300nm to 30nm). We  demonstrate a way of  extracting quantitative
information about the background impurity concentration in GaAs and AlGaAs, the interface roughness and the
charge in the surface states from the data. This information is very useful from the perspective of molecular beam epitaxy (MBE)
growth. It is difficult to fabricate such shallow high-mobility 2DEGs using modulation doping due to the
need to have a large enough spacer layer to reduce scattering and switching noise from remote ionsied dopants.
\end{abstract}

\pacs{73.40.Kp, 73.20.Mf} \keywords{undoped heterostructure, shallow ohmics, surface states} \maketitle
Identifying the sources of mobility limiting scattering  and relating them to growth conditions is an important practical problem in
semiconductor physics.  Minimising the background impurity concentration is crucial for achieving high mobility 2DEGs (and 2DHGs) as well as for
lowering the percolation threshold, below which the 2DEG breaks up into inhomogeneous puddles. It is also important in
reducing the probability of a gate defined nanostructure (like point contacts, quantum dots, interferometers) intercepting a defect/impurity.
In this context, a specific question,  that motivated this work is : how impure is the AlGaAs compared to GaAs? \\

The paper is organized in three parts. First we describe our  method for distinguishing the impurity concentrations
in GaAs and AlGaAs layers that form the heterointerface. We present experimental data from a series of wafers where the interface depth
was varied from 30 to 300 nm. Intimately connected with the success and applicability of this method is
a way of contacting very shallow (30nm) undoped 2DEGs. This forms the second part. Finally in conclusion we argue that this
method can have  ramifications in the fabrication of shallow nanostructures as well as studying the effect of surface states.\\

\begin{figure}
\includegraphics[width=0.48\textwidth,clip]{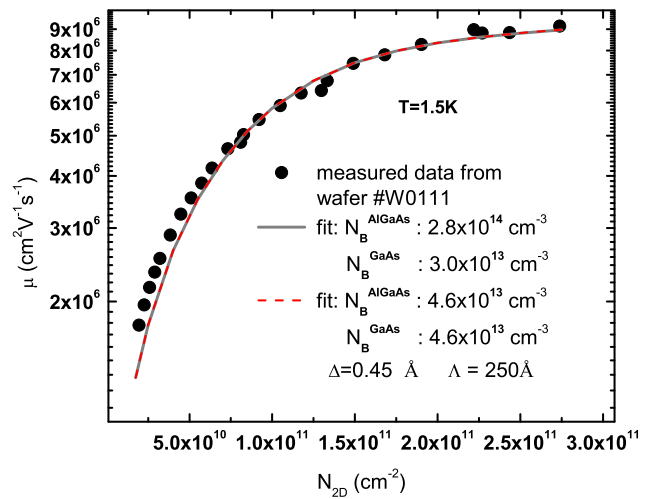}
\caption{\label{W0111data} (Colour online) Data from a high mobility wafer for which the MBE-chamber optimisation was carried out using the
process described in the paper. The 2DEG is 300nm deep and unaffected by the surface charges. Note however that two widely
different ratios (even if one is chosen to be unphysical) of  background impurity concentrations can give excellent fits to the experimental data. The dotted line and the continuous line are indistinguishable. It is this ambiguity that can be resolved by using a sequence of 3-4 wafers. In  cases where only one thickness of AlGaAs is available for some reason, we would set $N_{\rm B}^{\rm GaAs}$=$N_{\rm B}^{\rm AlGaAs}$.}
\end{figure}

\begin{figure}
\includegraphics[width=0.48\textwidth,clip]{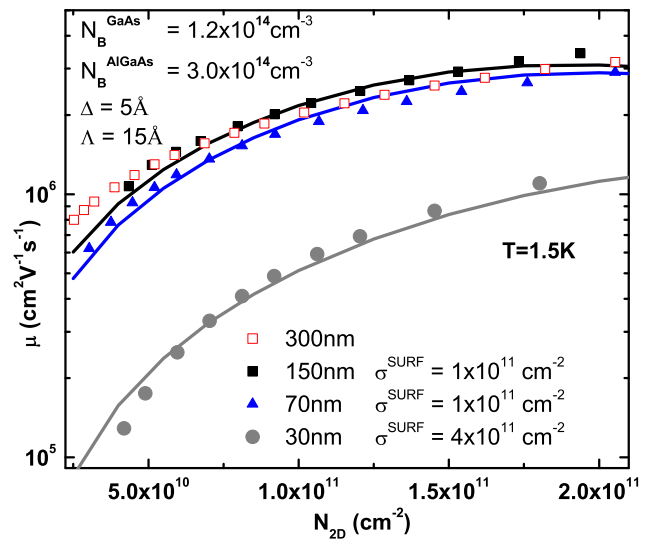}
\caption{\label{A25xxdata}(Colour online) Data from the four wafers (A2510,A2511,A2512,A2513) with d=30,70,150, 300 nm. The effect of the surface
states are clearly seen in the d=30nm sample.}
\end{figure}

Consider first the structure shown in Fig. \ref{schematic}. If  a 2DEG is
induced by a topgate $d\approx300{\rm nm}$ below the surface, at low
temperature (typically 1.5K or lower) the mobility limiting scattering would
come from two sources. First is the Coulomb scattering from the (charged)
impurities incorporated  during MBE growth. The effect of this is stronger at
lower densities and this background also determines a threshold below which the
2DEG would break up into inhomogenous  puddles. There are no modulation doped
or delta-doped layers, but the surface states have some bound charge and these
will affect the 2DEG if the AlGaAs layer is less than $d\sim 80{\rm nm}$ deep,
in a manner similar to a delta-doped layer. Secondly the GaAs-AlGaAs interface
is not perfectly smooth. The interface roughness can be  modelled  as a
perturbation, that affects the 2DEG more at higher densities than at lower
densities, giving rise to a scattering rate $\tau_{\rm IR}$.  Between the
percolation threshold and near complete occupancy of the first subband, the
effect of these two scattering processes can quantitatively explain the
observed variation of mobility with density (see Fig.\ref{W0111data} and eqn.
\ref{FormFactor_for_one_impurity_eqn}-\ref{muNeqn}). Observe however that the
scattering rate due to background impurities can be written as a sum of two
parts : $1/\tau_{\rm B}^{\rm AlGaAs}$ arising from impurities in AlGaAs
($N_{\rm B}^{\rm AlGaAs}$) and $1/\tau_{\rm B}^{\rm GaAs}$ arising from
impurities in GaAs ($N_{\rm B}^{\rm GaAs}$). From a single wafer we can  know
the sum of the two, but cannot determine how the impurities are distributed
between GaAs and AlGaAs.
Consider now a second wafer grown at the same time (in a sequence), in the same chamber but with a lesser thickness of
AlGaAs. The background and the interface roughness should remain the same. We then measure the density-mobility trace for the shallower wafer.
The contributions $\tau_{\rm B}^{\rm GaAs}$, $\tau_{\rm IR}$ can change only marginally. $1/\tau_{\rm B}^{\rm AlGaAs}$ must decrease because the integration over the thickness of the
AlGaAs now has a lesser span (eqn. \ref{AlGaAseqn}). If we assign a wrong value to $N_{\rm B}^{\rm AlGaAs}$ then the decrement in $1/\tau_{\rm B}^{\rm AlGaAs}$ would also be wrong. $1/\tau_{\delta}^{\rm SURF}$ should however increase in a predictable way, because the surface charges are now closer.  We grew a series of wafers (as in Fig. \ref{A25xxdata}) and fabricated heterostructure insulated gate field effect transistors (HIGFET) on each and modelled the $\mu(N)$ curves with the following constraints :

\begin{figure}
\includegraphics[width=0.4\textwidth,clip]{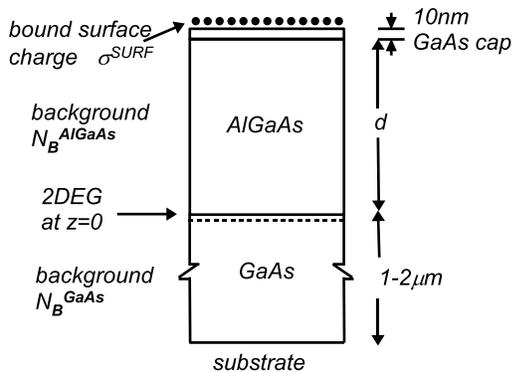}
\caption{\label{schematic} Schematic of the sample used in the experiments. Only the
thickness of the AlGaAs ($d$) varies between the wafers of the sequence. We have varied
$d$ from 20nm to 300nm. The effect of the bound charge in the surface states is considered
in exactly the same way as one treats scattering due to delta-doping layer. }
\end{figure}

\begin{enumerate}
\item{$N_{\rm B}^{\rm GaAs}$, $N_{\rm B}^{\rm AlGaAs}$  must remain the same for all the traces.}
\item{$\tau_{\rm IR}$ must remain the same for all the traces.}
\item{Surface states are modelled as  a $\delta$-doped layer at a known distance ($d+10{\rm nm}$) away.}
\end{enumerate}
To our knowledge there is no other method that distinguishes between the
impurity concentration in GaAs and AlGaAs. As Al is more reactive than Ga, with
impurities such as oxygen the background impurity levels in the two materials
can be expected to differ. Our results have consistently shown (for two MBE
chambers) that that $N_{\rm B}^{\rm AlGaAs}$ is about 2-3 times $N_{\rm B}^{\rm
GaAs}$. We can also determine quantitatively the amount of charge in the
surface states between GaAs and the insulator. It is important to note that
data from a sequence of 2-3 progressively shallower structures is crucial for
distinguishing the background levels in GaAs and AlGaAs.

Conventionally one can measure the background doping levels ($N_{\rm B}$) by growing a very thick layer
($t\sim$ few microns)  and measuring the areal carrier concentration ($N=N_{\rm B}t$). However
this method becomes harder as the background level drops to $\sim10^{14}{\rm cm}^{-3}$ and lower,
which is the typical level found in chambers capable of growing very high mobility structures.
Additionally to reach a regime where the measured carrier concentration per unit area ($N$) scales with thickness,
one has to first overcome the surface depletion regime. If the free surface takes up $10^{11} - 10^{12}{\rm cm}^{-2}$
carriers, then it is easy to see that to reach measurable areal densities, one would have to use a layer
at least 15-20 microns in thickness. Whereas in realistic wafers, the MBE grown region is typically not
more than 1-2 microns thick, including any buffer regions. Determining the impurity levels
in AlGaAs is harder, because the ohmic contacts to AlGaAs are more difficult and  the free surface of
AlGaAs oxidizes rapidly. Our transport based method can be used to determine the background levels in
both GaAs and AlGaAs which are grown as part of the same heterostructure.\\

The Boltzmann transport based formulation of scattering in a 2DEG\cite{Ando1982RMP,Gold1988PRB,Matsumoto1974JJAPS,Hirakawa1986PRB} has been used to explain the observed mobility of 2DEGs in
Si-MOSFETs and GaAs-AlGaAs heterointerfaces, with very good success. Briefly, one obtains the subband wavefunctions for the
interfacial 2DEG and then the form factors are obtained by integrating over the charge distribution of the lowest subband wavefunction
$\psi(z)$, following the usual procedure\cite{Ando1982RMP}
\begin{eqnarray}
\label{FormFactor_for_one_impurity_eqn}
F_{1}(q,z)&=&\int{dz'}|\psi(z)|^{2}e^{-q|z-z'|}\\
\label{FormFactor_for_wavefn_eqn}
F(q)&=&\int{dz}\int{dz'}|\psi(z)|^{2}|\psi(z')|^{2}e^{-q|z-z'|}
\end{eqnarray}
It is sufficient to consider the  dielectric screening in the Thomas-Fermi approximation for ($q<2k_{\rm F}$):
\begin{equation}
\epsilon(q)=1 + \frac{e^{2}}{2\epsilon_{0}\epsilon_{r}q}\frac{m}{\pi\hbar^2}F(q)
\end{equation}
yielding a scattering cross-section $\nu(\theta,z)$ due a charged impurity at a distance, $z$, from the interface
 \begin{equation}
 \nu(\theta,z)=\frac{me^4}{8\pi\hbar^3{\epsilon_0}^2{\epsilon_r}^2}\left|\frac{F_{1}(q,z)}{q\epsilon(q)}\right|^2
 \end{equation}
 where $q=2k_{\rm F}\sin\frac{\theta}{2}$. The scattering rates are then:
\begin{eqnarray}
\label{GaAseqn}
\frac{1}{\tau_{\rm B}^{\rm GaAs}} &=& N_{\rm B}^{\rm GaAs}\int^{0}_{-\infty}{dz} \int_{0}^{\pi}d\theta(1-\cos\theta)\nu(\theta,z)\\
\label{AlGaAseqn}
\frac{1}{\tau_{\rm B}^{\rm AlGaAs}} &=& N_{\rm B}^{\rm AlGaAs}{\int^{d}_{0}}{dz} \int_{0}^{\pi}d\theta(1-\cos\theta)\nu(\theta,z)\\
\label{SURFeqn}
\frac{1}{\tau_{\delta}^{\rm SURF}} &=&  \sigma^{\rm SURF}{\int_{0}^{\pi}}d\theta(1-\cos\theta)\nu(\theta,z^{\rm SURF})
\end{eqnarray}
The contribution of the interface roughness increases with increasing electron
concentration at the heterointerface \cite{Matsumoto1974JJAPS}.
 This (somewhat counter-intuitive)  result arises due to a faster growth of the electric field at the interface compared to
the intrinsic screening by the 2DEG itself which reduces the effect.
The roughness is described statistically by a correlation
\begin{equation}
\langle\Delta(r)\Delta(r')\rangle={\Delta^2}e^{-\frac{(r-r')^2}{\Lambda^2}}
\end{equation}
\begin{equation}
\label{IReqn}
\frac{1}{\tau_{\rm IR}} = \frac{m}{\hbar^3}(\Lambda\Delta)^2\int_{0}^{\pi}d\theta(1-\cos\theta)\left|\frac{e^2}{2{\epsilon_0}{}\epsilon_{r}}\frac{N}{2\epsilon(q)} \right|^2
e^{-\frac{{\Lambda^2}{q^2}}{4}}
\end{equation}

Finally we obtain
\begin{eqnarray}
\frac{1}{\tau^{\rm TOTAL}} &=& \frac{1}{\tau_{\rm B}^{\rm GaAs}}+ \frac{1}{\tau_{\rm B}^{\rm AlGaAs}} + \frac{1}{\tau_{\delta}^{\rm SURF}}+\frac{1}{\tau_{\rm IR}}\\
\label{muNeqn}
\mu(N) &=&\frac{e}{m}\tau^{\rm TOTAL}
\end{eqnarray}

Fig.\ref{A25xxdata} shows how the parameters can be extracted from a set of
measurements, satisfying the three consistency conditions mentioned earlier.
Note that for a 2DEG deeper than 100nm, the mobility hardly changes with depth.
This is fully consistent with the predictions of the analysis mentioned in the
preceding paragraph. The flattening of the curves at higher density is a
consequence of the interface roughness scattering, the lower density regions
are dominated by the effect of the charged background and in shallow devices
the charged surface states. The charge in the surface state is found to remain
constant over the entire range of gate voltage used. The typical density of the
surface charge that we obtain ($\sigma^{\rm SURF}\approx 1-5{\times}10^{11}{\rm
cm^{-2}}$) is considerably less than the typical charge in a delta doped layer
- typically in the $10^{12}{\rm cm^{-2}}$ range. The very shallow undoped 2DEG
not only offers a way to extract information about the wafer characteristics
but also considerable advantages over shallow doped wafers in terms of
mobility,
absence of parallel conduction and most likely much reduced switching noise. \\

Note that for a 2DEG, even at $n=1\times10^{10}{\rm cm}^{-2}$, $T/T_{F}\approx0.3$ at the temperature of measurement $T=1.5{\rm K}$
and we can consider the scattering at the Fermi surface only, hence no thermal (Boltzmann) averaging needs to be done to get the scattering rates. While we have also been able to induce a 2DHG in the same wafers (by using a negative top gate bias and p-type ohmic
contacts), the higher effective mass of the holes leads to a lower Fermi temperature, so the assumption $T/T_{F} << 1$ is no longer valid.
Larger correlations also make the simple calculation of $\epsilon(q)$ invalid.
Thus it remains much more convenient to use a 2DEG to extract the
background, surface charge and interface roughness parameters.

\begin{figure}
\includegraphics[width=0.48\textwidth,clip]{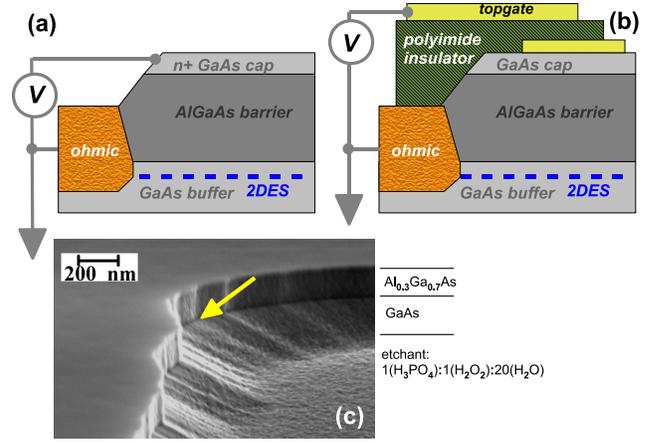}
\caption{\label{etchprofile}(Colour online) (a) Schematic of the undoped device with a doped topgate, (b) Insulated gate method where no
doped cap is used. (c) The etch profile found to be useful for good sidewall wetting.  Ni/AuGe/Ni was evaporated at an angle$\approx 60^{o}$ to the
perpendicular to the chip surface. The contacts were annealed at 470$^o$C for 120seconds in N$_2$/H$_2$ atmosphere.}
\end{figure}

\begin{figure}
\includegraphics[width=0.48\textwidth,clip]{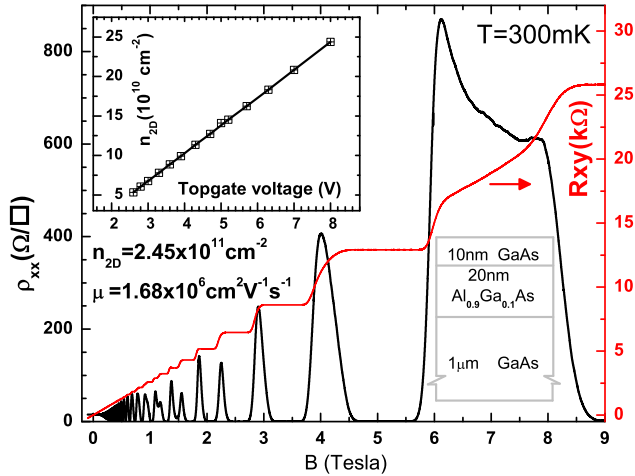}
\caption{\label{V627_20nm2DEG_data} (Colour online) Hall and Shubnikov-de Haas oscillations in a very shallow 2DEG, formed 30nm below the surface (wafer V627).
We have used a two level  gating scheme \cite{Harrell1999APL,Sarkozy2009APL}, one directly on the surface (see Fig. \ref{etchprofile}b) and one of top of the polyimide insulator
to define point contacts (data not shown). }
\end{figure}

The importance of shallow doped\cite{Laroche2010APL,Kozolov2007Semiconductors} and undoped 2DEGs/2DHGs \cite{Klochan2010APL,See2010APL}has recently been noted in the context of surface gate defined  nanostructures. However the method of making ohmic contacts to very shallow undoped 2DEGs requires some careful consideration. Fig. \ref{etchprofile}
a \cite{Kane1993APL,Klochan2010APL,See2010APL}\& \ref{etchprofile}b \cite{Harrell1999APL,Willett2006APL,Sarkozy2009APL,Sarkozy2009PRB} show the two known methods of making undoped 2DEGs.  However a structure of the type shown in Fig. \ref{etchprofile}a cannot be
made very shallow, because  the ohmic material (after annealing) is necessarily somewhat rough and tends to creep up the sidewalls
to some extent. In practice it is extremely difficult to prevent unwanted shorting paths between the ohmic and the doped cap that acts as a
topgate, if the AlGaAs layer is any less than 100nm thick. The structure of Fig. \ref{etchprofile}b does not suffer from this limitation
and can be made arbitrarily shallow. We have successfully fabricated devices where the AlGaAs is only 20nm thick with reliable ohmic contacts
working above ${\rm B}=10{\rm Tesla}$ and ${\rm T}<300{\rm mK}$. The ohmic material in these cases need to be evaporated into etched pits
(see Fig.\ref{etchprofile}c) at an angle of $60^{\rm o}$ to ensure good sidewall wetting. Experimentally we found (from AFM scans) that the
annealed ohmic near the AuGeNi-GaAs sidewall tends to be rough and spiky if the sidewalls are not fully wetted. Good diffusion into the sidewalls
 resulted in a comparatively smoother boundary ensuring good uniform coverage of the polyimide insulator (see Fig. \ref{etchprofile}b) between
 the ohmic and the overlapping topgate. The topgate to ohmic bias could be held at 10V with a leakage current of less than 50pA, with the
 resistance less than $50{\rm \Omega}$/contact (see Fig \ref{V627_20nm2DEG_data}).\\
 In conclusion, we have devised a low temperature transport based method of quantitatively distinguishing the  background impurities in GaAs and AlGaAs. We also determine that the amount of charge in the surface states is significantly less than what is expected in a typical delta
 doped layer. The method can be used to determine the backgrounds resulting from  growing a wafer at different growth rates, substrate temperatures and cell
 temperatures. As part of this process we developed a way of making good ohmic contacts to very shallow undoped heterostructures.
 In structures where modulation doping or delta doping is used a minimum distance from the surface is dictated by the need to have a large enough
spacer layer to reduce scattering and switching noise from remote ionised dopants. The undoped 2DEGs do not suffer from these limitations. Shallow, high mobility undoped 2DEGs are expected to be  useful in defining smaller lithographic features that conveniently approach the single electron limit.\\

The work was funded by EPSRC, UK. The authors thank C.A. Nicoll, J. Waldie and A.F. Croxall for useful suggestions and discussions.

\end{document}